\documentclass[aps,prl,twocolumn,floats]{revtex4}
\usepackage{epsfig}
\begin{document} 

\title{The Evolution of Magnetism in Iron from the Atom to the Bulk}

\author{Murilo L. Tiago$^{(a)}$, Yunkai Zhou$^{(b)}$
\footnote{Work done while at the Department of Computer Science \&
  Engineering, University of Minnesota, Minneapolis, MN 55455, USA.},
  M.M.G.\ Alemany$^{(c)}$, Yousef Saad$^{(d)}$, and James
  R. Chelikowsky$^{(a,e)}$} \affiliation{ $^{(a)}$ Center for
  Computational Materials, Institute for Computational Engineering and
  Sciences, University of Texas, Austin, TX 78712, USA. \\ $^{(b)}$
  Department of Mathematics, Southern Methodist University, Dallas TX
  75275, USA. \\ $^{(c)}$ Departamento de F\'{\i}sica de la Materia
  Condensada, Facultad de F\'{\i}sica, Universidad de Santiago de
  Compostela, E-15782 Santiago de Compostela, Spain \\ $^{(d)}$
  Department of Computer Science \& Engineering, University of
  Minnesota, Minneapolis, MN 55455, USA.\\ $^{(e)}$ Departments of
  Physics and Chemical Engineering, University of Texas, Austin, TX
  78712, USA.}

\date{\today}

\begin{abstract}
The evolution of the magnetic moment in iron clusters containing 20 to
400 atoms is investigated using first-principles numerical
calculations based on density-functional theory and real-space
pseudopotentials. Three families of clusters are studied,
characterized by the arrangement of atoms: icosahedral, body-centered
cubic centered on an atom site, and body-centered cubic centered on
the bridge between two neighboring atoms. We find an overall decrease
of magnetic moment as the clusters grow in size towards the bulk
limit. Clusters with faceted surfaces are predicted to have magnetic
moment lower than other clusters with similar size. As a result, the
magnetic moment is observed to decrease as function of size in a
non-monotonic manner, which explains measurements performed at low
temperatures.
\end{abstract}

\maketitle


The existence of spontaneous magnetization in metallic systems is an
intriguing problem because of the extensive technological applications
of magnetic phenomena and an incomplete theory of its fundamental
mechanisms. Clusters of metallic atoms are important in this respect
as they serve as a bridge between the atomic limit and the bulk, and
can form a basis for understanding the emergence of magnetization as
function of size. Several phenomena such as ferromagnetism, metallic
behavior, and ferroelectricity have been intensely explored in bulk
metals, but the way they manifest themselves in clusters is an open
topic of debate.  At the atomic level, ferromagnetism is associated
with partially filled $3d$ orbitals. In solids, ferromagnetism may be
understood in terms of the itinerant electron model \cite{mattis},
which assumes partial delocalization of the $3d$ orbitals. In clusters
of iron atoms, delocalization is weaker owing to the presence of a
surface, whose shape affects the magnetic properties of the
cluster. Because of their small size, iron clusters containing a few
tens to hundreds of atoms are superparamagnetic: the entire cluster
serves as a single magnetic domain, with no internal grain boundaries
\cite{bean59}.  Consequently, these clusters have strong magnetic
moments, but exhibit no hysteresis.

The magnetic moment of nano-sized clusters has been measured as
function of temperature and size \cite{billas93,gerion00,billas94},
and several aspects of the experiment have not been fully clarified,
despite the intense work on the subject
\cite{franco99,postnikov04,sipr04,postnikov04-1,felix-medina03,edmonds99}. One
intriguing experimental observation is that the specific heat of such
clusters is lower than the Dulong-Petit value, which may be due to a
magnetic phase transition \cite{gerion00}. In addition, the magnetic
moment per atom does not decay monotonically as function of the number
of atoms and for fixed temperature. Possible explanations for this
behavior are: structural phase transitions, strong dependence of
magnetization with the shape of the cluster, or coupling with
vibrational modes \cite{gerion00}.  One difficulty is that the
structure of such clusters is not well known. First-principles and
model calculations have shown that clusters with up to 10 or 20 atoms
assume a variety of exotic shapes in their lowest-energy configuration
\cite{pastor89,dieguez01}. For larger clusters, there is evidence for
a stable body-centered cubic (BCC) structure, which is identical to
ferromagnetic bulk iron \cite{franco99}.

The evolution of magnetic moment as function of cluster size has
attracted considerable attention
\cite{billas94,billas93,gerion00,franco99,postnikov04,sipr04,pastor89,postnikov04-1,felix-medina03,edmonds99,dieguez01}.
A key question to be resolved is: What drives the suppression of
magnetic moment as clusters grow in size?  In the iron atom, the
permanent magnetic moment arises from exchange splitting: the
$3d_\uparrow$ orbitals (majority spin) are lower in energy and
completely occupied with 5 electrons, while the $3d_\downarrow$
orbitals (minority spin) are partially occupied with one electron,
resulting in a magnetic moment of $4 \mu_B$, $\mu_B$ being the Bohr
magneton. When atoms are assembled in a crystal, atomic orbitals
hybridize and form energy bands: $4s$ orbitals create a wide band
which remains partially filled, in contrast with the completely filled
$4s$ orbital in the atom; while the $3d_\downarrow$ and $3d_\uparrow$
orbitals create narrower bands. Orbital hybridization together with
the different bandwidths of the various $3d$ and $4s$ bands result in
weaker magnetization, equivalent to 2.2 $\mu_B/$atom in bulk iron.

\begin{figure}[h]
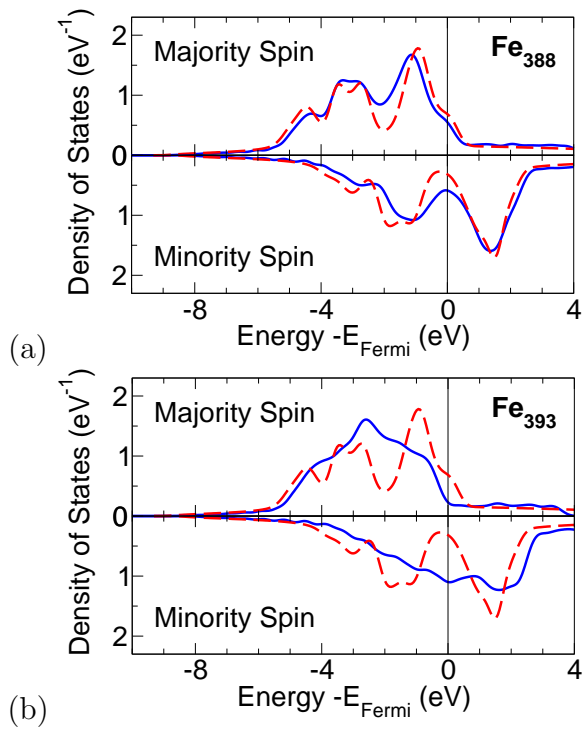

{\large (a)} \epsfig{figure=Fig1a.eps,width=7cm,clip=}

{\large (b)} \epsfig{figure=Fig1b.eps,width=7cm,clip=}
\caption{Density of states in the clusters
Fe$_{388}$(a) and Fe$_{393}$(b),
majority spin (upper panel) and minority spin(lower panel). Fe$_{388}$
corresponds to a fragment of the BCC crystal. Fe$_{393}$ has local
icosahedral coordination. For
reference, the density of states in bulk iron is shown in dashed
lines. The Fermi energy is chosen as energy reference.}
\label{f_dos}
\end{figure}

In atomic clusters, orbital hybridization is not as strong because
atoms on the surface of the cluster have fewer neighbors. The strength
of hybridization can be quantified by the effective coordination
number. A theoretical analysis of magnetization in clusters and thin
slabs indicates that the dependence of the magnetic moment with the
effective coordination number is approximately linear
\cite{sipr04,felix-medina03,edmonds99}. But the suppression of
magnetic moment from orbital hybridization is not isotropic
\cite{bruno93}.  For instance, if we consider a layer of atoms on the
$x-y$ plane, the $3d$ orbitals oriented on the plane (the ones with
angular dependence $xy$ and $x^2-y^2$) will hybridize more effectively
than orbitals oriented normal to the plane (with angular dependence
$3z^2-r^2$), because of increased overlap among orbitals on the $x-y$
plane. If the layer has infinite extent, orbitals oriented on that
plane will tend to form an energy band wider than the one formed by
orbitals normal to it. With a wider band, the difference in electron
population between the majority spin channel and the minority spin
channel is reduced, leading to reduced spin polarization and overall
weaker magnetization. Reduced orbital hybridization has been observed
to enhance the magnetic moment of metallic clusters \cite{bansmann05}
and thin layers \cite{freeman91}.  As a consequence of this
anisotropy, clusters with faceted surfaces are expected to have
magnetic properties different from clusters with irregular surfaces,
even if they have the same effective coordination number.  This effect
is likely responsible for a non-monotonic suppression of magnetic
moment as function of cluster size. In order to analyze the role of
surface faceting, we have performed first-principles calculations of
the magnetic moment of iron clusters with various geometries and with
sizes ranging from 20 to 400 atoms.

We determine the electronic structure of clusters within the framework
of pseudoptentials\cite{troullier91} constructed using
density-functional theory (DFT) \cite{martin,parr,payne92}. DFT is  an
established theory for first-principles studies of weakly and
moderately correlated electronic systems.  The exchange-correlation
functional used in this work employs the generalized gradient
approximation (GGA) \cite{perdew96}. We have observed that the GGA
predicts magnetic moments enhanced with respect to the simpler
local-density approximation by 2~\% to 10~\%.  For a fixed geometry of
the cluster, we solve self-consistently the Kohn-Sham equation on a
regular grid in real space \cite{chelikowsky94,chelikowsky94_a}.
Proper boundary conditions are obeyed by imposing the electronic wave
functions to vanish on the boundary of a large spherical domain, which
contains the system of interest.  No explicit basis set is
used. Numerical convergence is controlled with two parameters: the
radius of the domain (typically 5 $\AA$ larger than the cluster
radius) and the nearest-neighbor spacing in the regular grid.  For
iron atoms, we use a spacing of 0.3 $a.u.$, (approximately 0.16 $\AA$).

We use the PARSEC code\cite{chelikowsky94,chelikowsky94_a}. This code
makes use of symmetry properties of the system and very efficient
techniques for solving the Kohn-Sham
equation\cite{zhou06,zhou06_a}. Until recently, a significant fraction
of numerical effort was spent in performing exact diagonalization of
the Kohn-Sham equation
\cite{chelikowsky94,chelikowsky94_a,payne92,beck00}. Curently, this
step is replaced with a series of subspace filtering iterations with
Chebyshev polynomials, which reduce the overall numerical effort by
one order of magnitude or more\cite{zhou06}. This dramatic advance in
methodology allows us to study confined systems with hundreds, if not
thousands of atoms in a straightforward and computationally efficient
manner.

\begin{figure}[h]
\centering\epsfig{figure=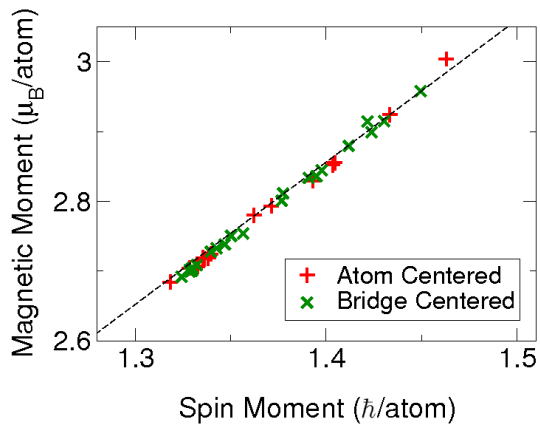,width=7cm,clip=}
\caption{Magnetic moment versus spin moment
  calculated for the atom-centered BCC (``plus'' signs) and bridge-centered
  BCC (crosses) iron clusters. The approximate ratio is $M/{\langle
  S_z \rangle} = g_{eff} = 2.04 \mu_B/\hbar$.}
\label{f_geff}
\end{figure}

Clusters of both icosahedral and BCC symmetry are explored in our
work. In order to investigate the role of surface faceting, we
construct clusters with faceted and non-faceted surfaces. Faceted
clusters are constructed by adding successive atomic layers around a
nucleation point. Small faceted icosahedral clusters exist with sizes
13, 55, 147, and 309. Faceted BCC clusters are constructed with BCC
local coordination and, differently from icosahedral ones, they do not
need to be centered on an atom site. We consider two families of cubic
clusters: atom centered or bridge centered, respectively for clusters
with nucleation point at an atom site or on the bridge between two
neighboring atoms. The lattice parameter is equal to the bulk value,
2.87 $\AA$.  Non-faceted clusters are built by adding shells of atoms
around a nucleation point so that their distance to the nucleation
point is less than a specified value. As a result, non-faceted
clusters usually have narrow steps over otherwise planar surfaces and
the overall shape is almost spherical.  By construction, non-faceted
clusters have well-defined point-group symmetries: $I_h$ or $T_{h}$
for the icosahedral family; $O_h$ for the atom-centered family; and
$D_{4h}$ for the bridge-centered family. Clusters constructed in that
manner show low tension on the surface, making surface reconstruction
less likely.  Our calculations indicate that atoms on the surface feel
forces weak in magnitude and directed towards the center of the
cluster. Owing to the small surface/volume ratio in large clusters,
the impact of surface reconstruction on those clusters will be small,
if not negligible.

As clusters grow in size, their properties approach the properties of
bulk iron. Figure \ref{f_dos}(a) shows the density of states (DOS) for
Fe$_{388}$, with local BCC coordination. At this size range, the
density of states assumes a shape typical of bulk iron, with a
three-fold partition of the $3d$ bands. In addition, the cohesive
energy of this cluster is only 77 meV lower than in bulk. This
evidence suggests that interesting size effects will be predominantly
observed in clusters smaller than Fe$_{388}$.  Figure \ref{f_dos}(b)
shows the DOS for Fe$_{393}$, which belongs to the icosahedral
family. This cluster has a very smooth DOS, with not much structure
compared to Fe$_{388}$ and bulk BCC iron. This is due to the
icosahedral-like arrangement of atoms in Fe$_{393}$. The overall
dispersion of the $3d$ peak ( 4~eV for $3d_\uparrow$ and 6~eV for
$3d_\downarrow$) is nevertheless similar in all the calculated DOS.

\begin{figure}[h]
\centering\epsfig{figure=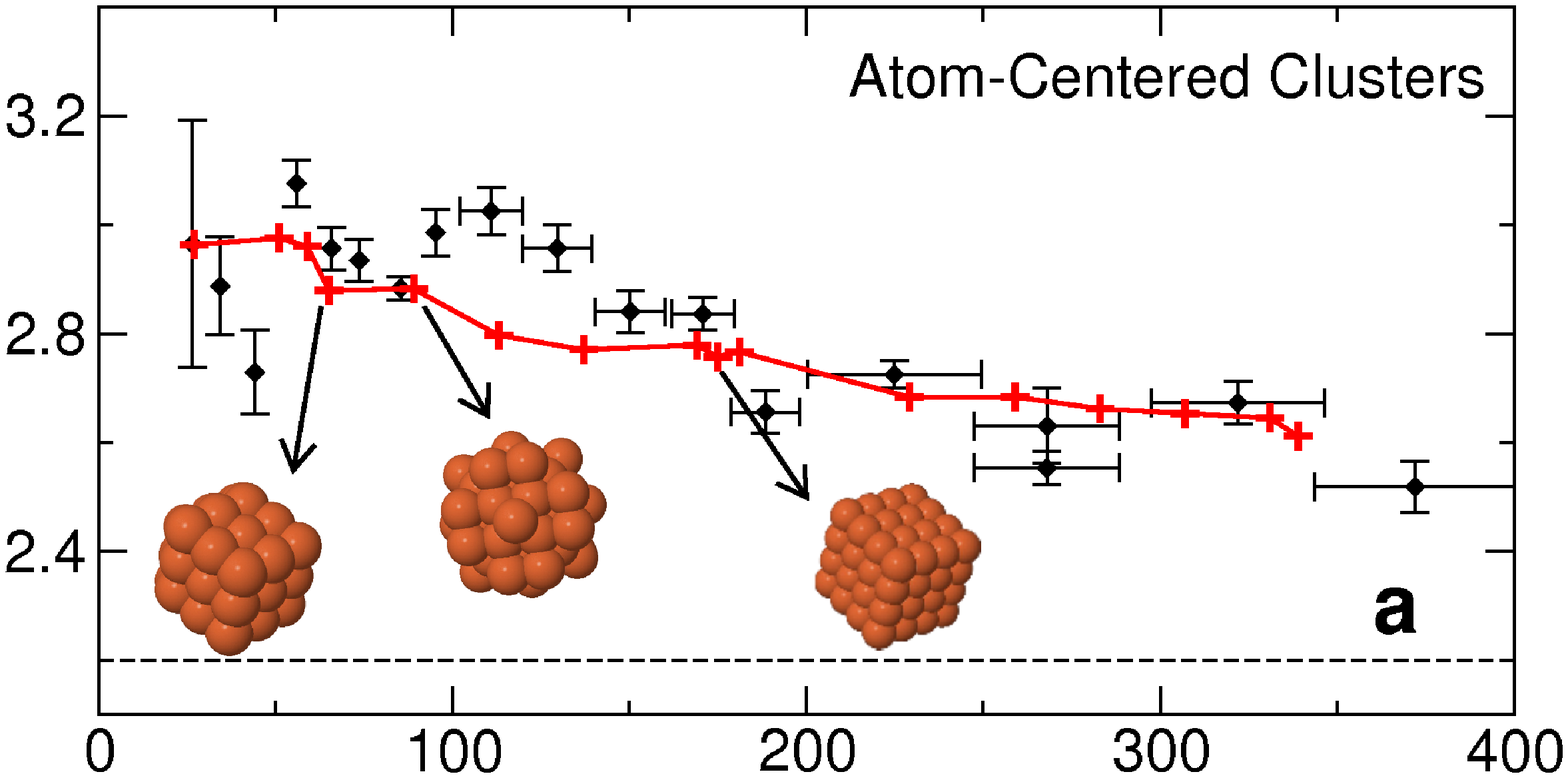,width=8cm,clip=}
\centering\epsfig{figure=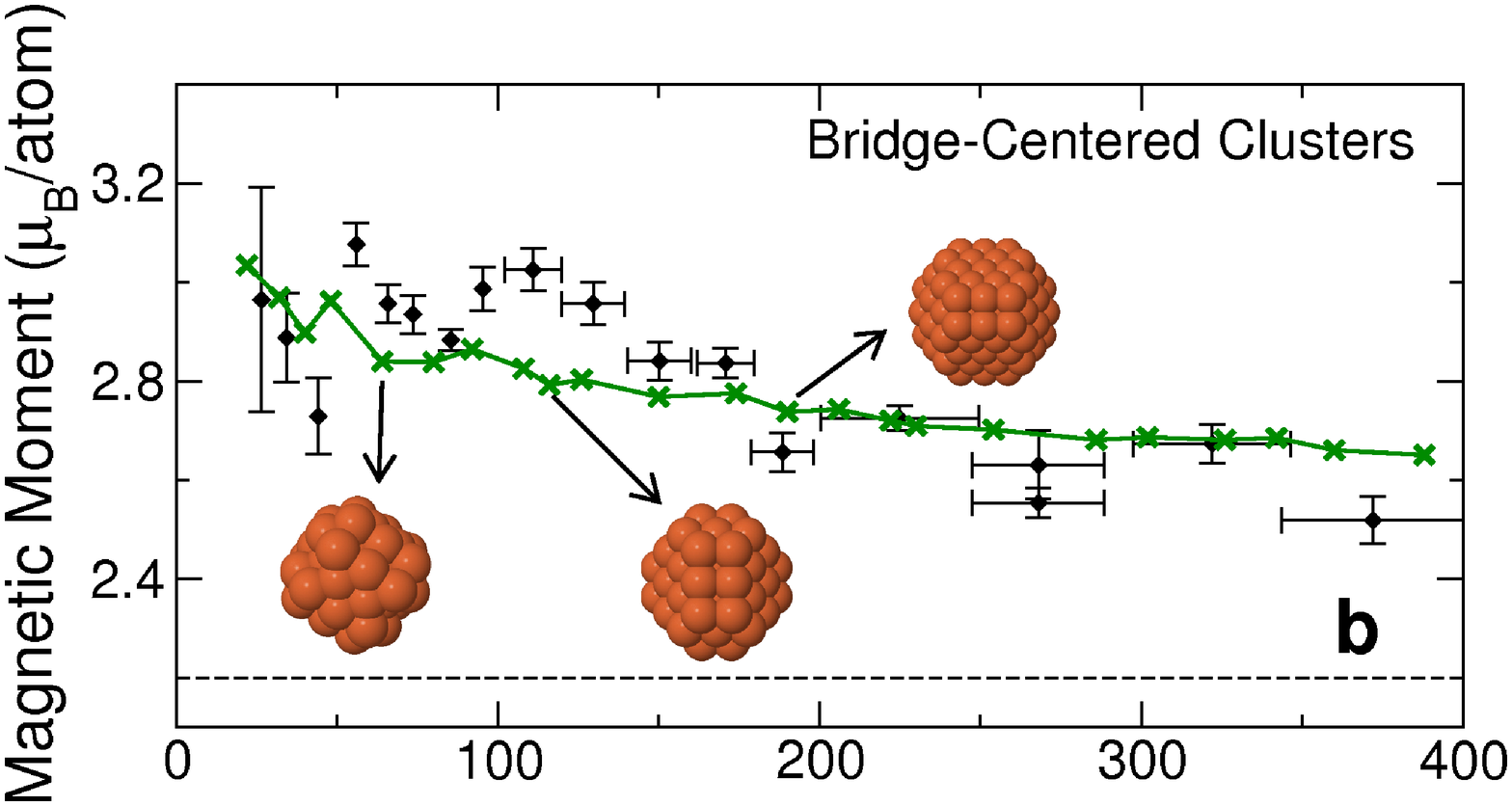,width=8cm,clip=}
\vspace{0.1cm}
\centering\epsfig{figure=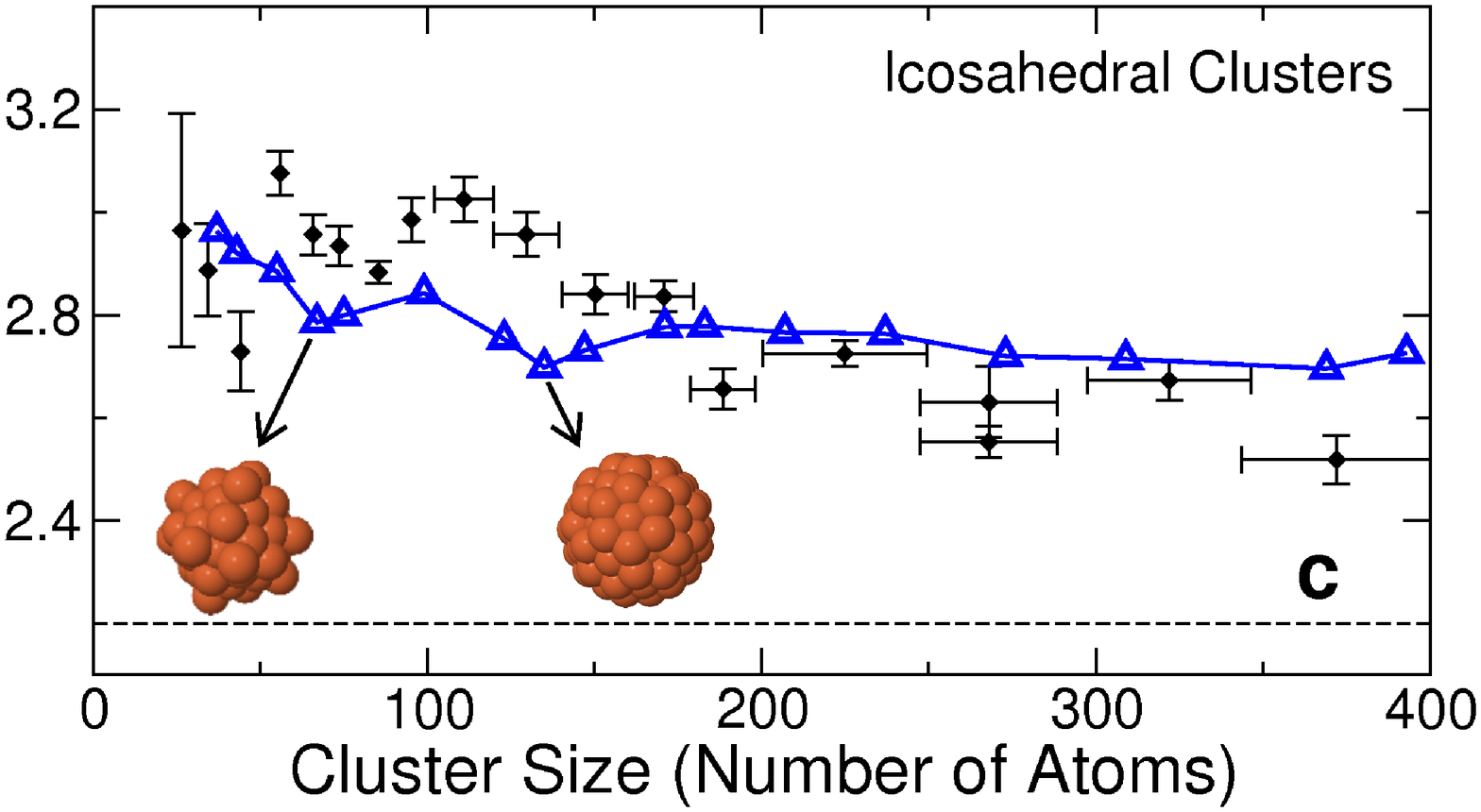,width=8cm,clip=}
\caption{Calculated magnetic moments for clusters in
  the atom-centered (``plus'' signs, a), bridge-centered (crosses, b),
  and icosahedral (triangles, c) families. Experimental data
  \cite{billas93} is shown in black diamonds with error bars. Some of
  the faceted and non-faceted clusters are depicted next to their
  corresponding data points. The dashed lines indicate the value of
  magnetic moment per atom in bulk iron.}
\label{f_magn}
\end{figure}

The magnetic moment is calculated as the expectation value of the
total angular momentum:

\begin{equation}
M = \frac {\mu_B}{ \hbar} \left[ g_s \left\langle S_z  \right\rangle
+\left\langle L_z \right\rangle  \right] =  \mu_B \left[ {g_s \over 2}
\left( n_\uparrow -n_\downarrow \right) +  {1 \over \hbar}
\left\langle L_z \right\rangle \right]
\label{e.magnetization}
\end{equation}
where $g_s = 2$ is the electron gyromagnetic ratio. Figure
 \ref{f_geff} illustrates the approximately linear dependence between
 the magnetic moment and spin moment, $\left\langle S_z
 \right\rangle$, throughout the whole size range. This results in an
 effective gyromagnetic ratio $g_{eff} = 2.04 \mu_B/\hbar$, which is
 somewhat smaller than the gyromagnetic ratio in bulk BCC iron, $2.09
 \mu_B/\hbar$. The difference in ratios is probably due to an
 underestimation in the orbital contribution, $\left\langle L_z
 \right\rangle $. In the absence of an external magnetic field,
 orbital magnetization arises from the spin-orbit interaction, which
 is included in the theory as a model potential,

\begin{equation}
V_{so} = - \xi {\bf L} \cdot {\bf S}
\label{e.so}
\end{equation}
where $\xi$ = 80 meV/$\hbar^2$ \cite{sipr04}.

Figure \ref{f_magn} shows the magnetic moment of several clusters
belonging to the three families studied: atom-centered BCC (top
panel), bridge-centered BCC (middle panel) and icosahedral (bottom
panel). Experimental data obtained  by Billas and collaborators
\cite{billas93} is also shown. The suppression of magnetic moment as
function of size is readily observed. Also, clusters with faceted
surfaces are predicted to have magnetic moments lower than other
clusters with similar sizes. This is attributed to more effective
hybridization of $d$ orbitals along the plane of the facets.  We also
notice that the correlation between magnetic moment and surface
smoothness is not always well defined. For instance, the cluster
Fe$_{175}$ in Figure \ref{f_magn}(a) has wide facets but its magnetic
moment is not much weaker than clusters with similar sizes. In Figure
\ref{f_magn}(c), clusters with strongly suppressed magnetic moment
have 67 and 135 atoms, whereas clusters with wide facets and similar
sizes have 55 and 147 atoms.

The measured non-monotonic behavior of magnetic moment can be
attributed to the shape of the surface. Under this assumption, islands
of low magnetic moment (observed at sizes 45, 85 and 188) are
associated to clusters with faceted surfaces. In the icosahedral
family, the islands of low magnetic moment are located around faceted
clusters containing 55, 147, and 309 atoms. The first island is
displaced by 10 units from the measured location. For the
atom-centered and bridge-centered families, we found islands at (65,
175) and (92, 173) respectively, as indicated in Figure
\ref{f_magn}(a,b).  The first two islands are also close to the
measured islands at 85 and 188. Clearly, there is no exact
superposition in the location of calculated islands and measured
islands. The magnetic moment was measured in clusters at 120 K
\cite{billas93,billas94}. At that temperature, vibrational modes or
the occurrence of metastable configurations can shift the islands of
low magnetic moment or make them more diffuse. Assuming that the
non-monotonic decay of magnetic moment is dictated by the cluster
shape, we also conclude that clusters with local structures different
from the ones we discuss here (such as cobalt clusters with
hexagonal-close packed coordination, or nickel clusters with
face-centered cubic coordination) should have islands of low magnetic
moment located at different ``magic numbers'', according to the local
atomic coordination.

In summary, we discussed the behavior of magnetization in iron
clusters containing 20 to 400 atoms in the light of first-principles
density-functional theory. The magnetic moment is found to decay as
function of cluster size in a non-monotonic fashion: clusters with
faceted surfaces are predicted to have magnetic moments lower than
clusters with non-faceted surfaces. As a consequence, clusters with
many steps, or atoms protruding from the surface, are expected to have
strong magnetic properties at low temperatures. In addition, large
clusters with icosahedral structure are expected to have magnetic
moments lower than clusters with BCC structure.

This work was supported by the National Science Foundation under
DMR-0130395 and DMR-0551195 and by the U.S. Department of Energy under
DE-FG02-89ER45391 and DE-FG02-03ER15491. Calculations were performed
at the Minnesota Supercomputing Institute (MSI), at the National
Energy Research Scientific Computing Center (NERSC), and at the Texas
Advanced Computing Center (TACC). M.M.G.A. acknowledges support from
the Spanish Ministry of Education and Science (Program ``Ram\'on y
Cajal'').

\end{document}